%% file: ICRC2023_template_IceCube.tex
\documentclass[a4paper,11pt]{article}

\usepackage{pos}
\usepackage{lipsum} 
\usepackage{subcaption}
\usepackage{siunitx}
\usepackage{pgf}
\usepackage{wrapfig}
\usepackage{svg}
\usepackage{todonotes}
\title{Three-year performance of the IceAct telescopes at the IceCube Neutrino Observatory}

\ShortTitle{IceAct: Three-year performance}

\author{The IceCube Collaboration \\{\normalsize \normalfont(a complete list of authors can be found at the end of the proceedings)}\\}

\emailAdd{lars.heuermann@rwth-aachen.de}

\abstract{

IceAct is an array of compact Imaging Air Cherenkov Telescopes at the ice surface as part of the IceCube Neutrino Observatory. The telescopes, featuring a camera of 61 silicon photomultipliers and fresnel-lens-based optics, are optimized to be operated in harsh environmental conditions, such as at the South Pole. Since 2019, the first two telescopes have been operating in a stereoscopic configuration in the center of IceCube's surface detector IceTop. With an energy threshold of about \SI{10}{TeV} and a wide field-of-view, the IceAct telescopes show promising capabilities of improving current cosmic-ray composition studies: measuring the Cherenkov light emissions in the atmosphere adds new information about the shower development not accessible with the current detectors. First simulations indicate that the added information of a single telescope leads, e.g., to an improved discrimination between flux contributions from different primary particle species in the sensitive energy range.
We review the performance and detector operations of the telescopes during the past 3 years (2020-2022) and give an outlook on the future of IceAct.

\vspace{4mm}
{\bfseries Corresponding authors:}
Lars Heuermann$^{1*}$\\
{$^{1}$ \itshape III. Physikalisches Institut, RWTH Aachen University, D-52056 Aachen, Germany}\\[4mm]
$^*$ Presenter

\ConferenceLogo{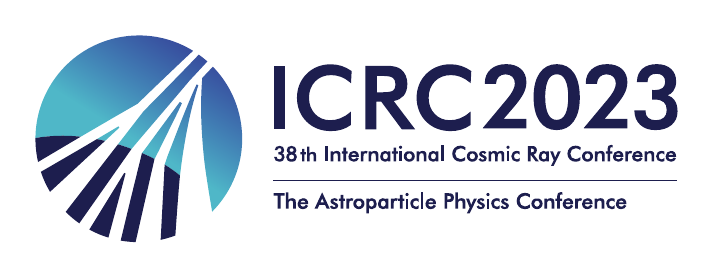}

\FullConference{The 38th International Cosmic Ray Conference (ICRC2023)\\ 26 July -- 3 August, 2023\\ Nagoya, Japan}
}

\begin{document}

\maketitle

\section{Introduction}\label{introduction}

The IceCube Neutrino Observatory \cite{Aartsen:2016nxy} is a cubic-kilometer-sized high-energy neutrino detector.
While its primary purpose is the detection of galactic and extra-galactic neutrinos, it is furthermore a unique cosmic ray detector.
Together with its surface array IceTop \cite{icetop} measurements of the Cosmic Ray spectrum with a respective measurement of the elemental composition \cite{composition_paper} have been done.
\begin{wrapfigure}{r}{0.45\textwidth}
    \centering
    \includegraphics[width=0.45\textwidth]{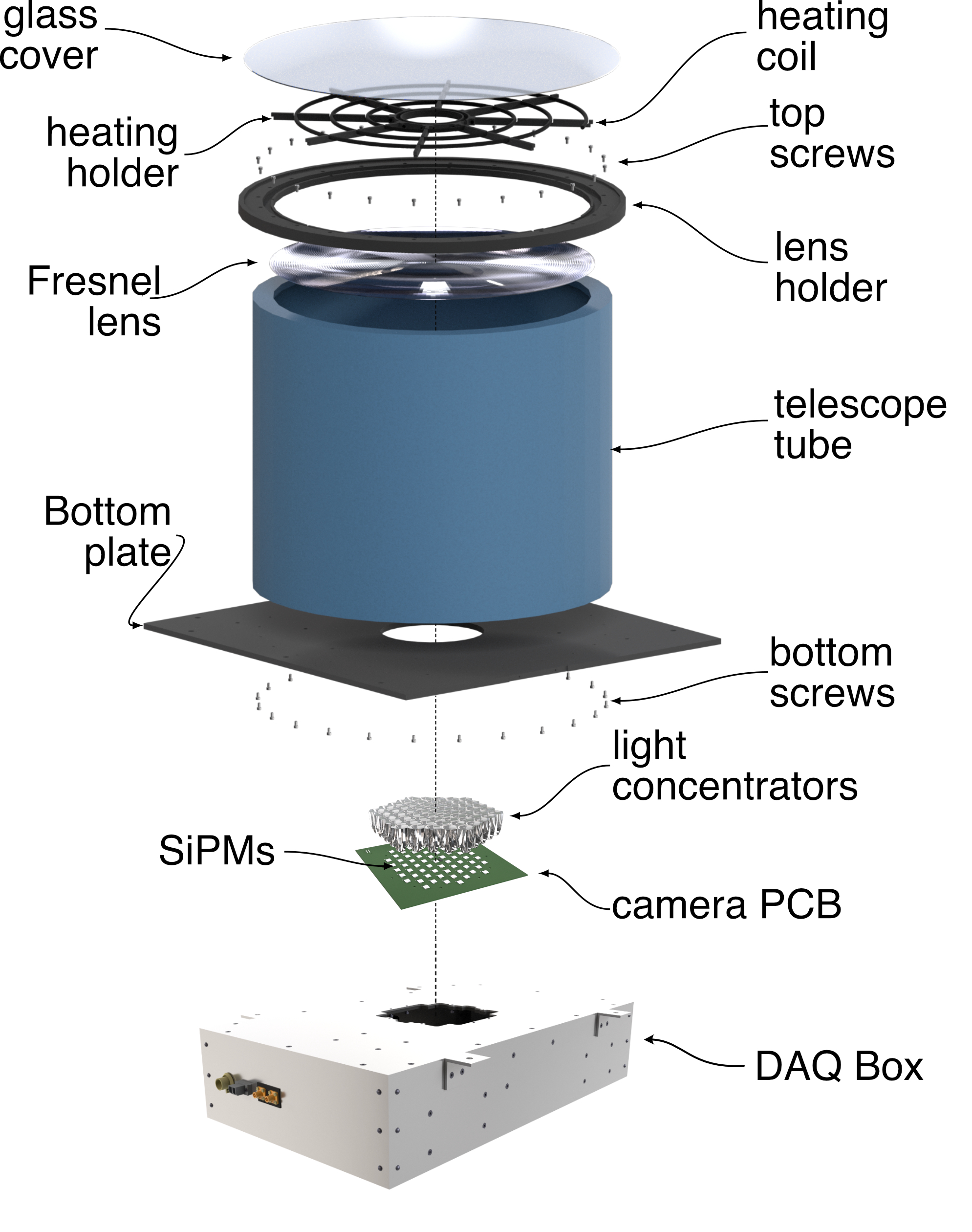}
    \caption{Explosion view of an IceAct telescope (not to scale). The camera is mounted inside the DAQ-box, the light concentrators fit through the holes in the bottom plate and the DAQ Box}
    \label{fig:exploded_IceAct}
\end{wrapfigure}
The main detector of IceCube \cite{Aartsen:2016nxy} (here referred to as \emph{in-ice}) consists of 5160 so-called digital optical modules (DOMs), each with a \SI{10}{inch} photo-multiplier-tube together with signal capture and readout electronics. The DOMs are installed in depths between \SI{1.45}{km} and \SI{2.45}{km}, arranged in 86 horizontal strings, resulting in 60 DOMs per string. The DOMs measure the emitted Cherenkov light of relativistic charged particles in the ice.
The IceTop detector consists of 162 tanks filled with ice, also detecting Cherenkov light.

IceAct is a proposed array of imaging air Cherenkov telescopes (IACTs) on the surface above the in-ice detector.
The primary purpose of IceAct is to extend the capabilities of the surface detector into the TeV region of primary cosmic ray energy.
It enables \cite{Schaufel:2019aef, IceActICRC2021} a hybrid measurement of cosmic rays. This opens possibilities for a cross-calibration of the energy threshold and energy scale of IceTop,
improving the mass measurement for a composition analysis of cosmic rays, and, potentially, 
improving the veto capabilities for atmospheric neutrinos in astrophysical neutrino measurements.

\section{Detector Set-Up}
\label{sec:setup}

The IceAct air Cherenkov telescopes (\autoref{fig:exploded_IceAct}) are based on the design described in \cite{Bretz:2018lhg, IceCube:2019yev}. 
Currently, two telescopes operate at the South Pole above the IceCube neutrino observatory. One is located on the roof of the IceCube Laboratory (referred to as \emph{roof}) and one (referred to as \emph{field}) is located \SI{220}{m} west of the roof on an aluminum stand close to the surface enhancement prototype station \cite{shefali}.

Each telescope features a 55-cm-diameter Fresnel lens and a 61-pixel camera with \qtyproduct{6x6}{mm} silicon photomultipliers (SiPMs) of the type MicroFJ-60035 produced by onsemi. The telescope tube is made out of fiberglass. The lens itself is shielded by a thin glass plate from the antarctic environment.
A self-regulating heating element is installed between the glass plate and the Fresnel lens to prevent and remove snow accumulations on the glass plate.
On each SiPM a Winston-cone-like light concentrator (referred to as \emph{cone}) out of Polymethylmethacrylate (PMMA) is glued. The top of each cone is shaped hexagonally, while the bottom is square with the same dimensions as a SiPM. The shape is optimized for maximal light yield while minimizing the dead area on the camera board.
The resulting field of view is roughly \SI{12}{\degree}, with a resolution of \SI{1.5}{\degree} per pixel.


Since 2020, both telescopes are equipped with a data acquisition (DAQ) based on the \mbox{TARGET-C} \cite{Leach:2020vht} application-specific integrated circuit. These modules (referred to as \emph{Targets}) were initially developed for the Compact High Energy Camera (CHEC) of the Cherenkov Telescope Array and are used with only minor modifications.
The most important features are the 64-channel readout and digitization with a \SI{1}{GS/s} sampling rate, the \SI{4096}{ns} deep storage capacitor array for each channel, the adjustable trigger threshold and supply voltage, the pile-up reducing signal shaping circuit - as well as the analog integration and digitization circuit referred to as \emph{slow analog-to-digital converter} (SADC).
The signal of four SiPMs is summed up, the trigger decision is always based on the sum - resulting in 16 so-called \emph{trigger groups}. The adjustable bias voltage is shared between the SiPMs of a trigger group. The storage capacitor array is set up as a continuously sampling ring buffer.
The modifications for usage in IceAct in comparison to CHEC are minor:
The readout of the buffer is increased to \SI{256}{ns} in comparison to the \SI{96}{ns} stated in \cite{Leach:2020vht}, the components of the shaping circuit are optimized for a different type of SiPM and the voltage range of the supply is adjusted to fit the operational range of the SiPM.

\section{Operations}\label{sec:ops}
\begin{figure}[t]
    \centering
    \resizebox{0.9\textwidth}{!}{\input{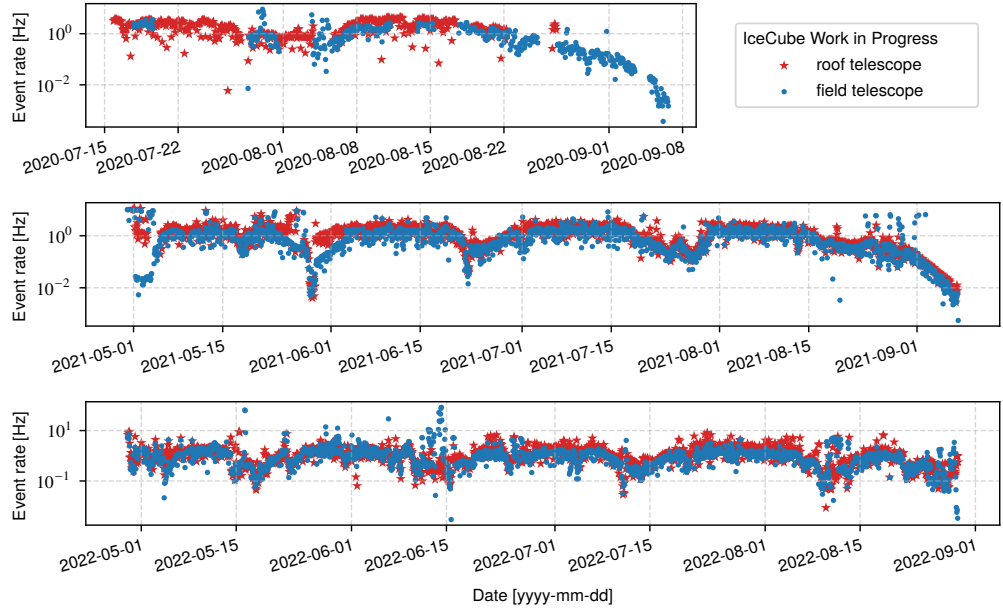}}
    \caption{90-min-averaged event rate for the roof and field telescopes during the three years of operation of IceAct in the current configuration. Each row corresponds to a different data-taking season, split by years.}
    \label{fig:EventRate}
\end{figure}

As of June 2023, the IceAct telescopes are in their fourth data-taking season with the current configuration.
Each season is structured the same way: Two types of calibration runs, the so-called muon run and the so-called dark noise run are taken (see \autoref{sec:Calib}) with both telescopes covered, i.e. closed in regards to ambient light from the moon, stars and twilight.
With the end of the astronomical twilight in May, the telescopes are uncovered and the data-taking is started.
The data-taking ends with the beginning of twilight in August. A firmware bug in the field programmable gate array in 2020 delayed the start to July. The start and end of the data-taking of each season can also be seen in \autoref{fig:EventRate}.

Every 6 hours the trigger threshold is adjusted with a full trigger scan (see \autoref{sec:Calib}) and a pedestal run (see also \autoref{sec:Calib}) is taken.
During 2021 the roof telescope took a trigger scan every hour and no pedestal runs.
The pedestal correction for 2021 is done by using a pedestal run from another year with the best matching temperature.
A readout of the ring buffer is triggered by next-neighbor-coincidence of two neighboring trigger groups and is considered an event.
The event rate is limited by a dead time to prevent a denial-of-service of the DAQ in the case of high event rates.
The temperature-dependent gain of the SiPMs is stabilized by measuring the temperature of the cameras and adjusting the bias voltage accordingly. For that purpose lookup tables for each trigger group have been produced before operations. In the case of strong ambient light the voltage and thus the gain of the SiPMs is reduced.
The resulting event rates for each season can be seen in \autoref{fig:EventRate}.
The event rate is stable around a few Hz during most of the season. Higher night-sky background (NSB, see \autoref{sec:Calib}) caused by the moon leads to an increase of the trigger threshold, causing a lower rate, which can be seen as periodic dips in the event rate.

Every 10 minutes operational parameters are logged. These include the temperature of the Target, the individual bias voltages of each trigger group, the SADC readout for each SiPM (see \autoref{sec:SADC}), the temperature of the camera board, as well as the event rate averaged over the last 100 events in the software buffer.
During operations pictures from the sky above the detector are taken with an \emph{all-sky cam} with three different exposure times every ten minutes.

\section{Calibration runs} \label{sec:Calib}


\paragraph*{Trigger scans.}
As IACTs are optical instruments monitoring the night-sky, they also measure ambient light. This is referred to as night-sky background (NSB). The amount of ambient light can change within minutes, especially during twilight \cite{Biland:2014fqa}.
The change of ambient light makes it necessary to frequently re-adjust the threshold for the trigger decision.
In the case of IceAct, the ambient light is more stable due to the longevity of the polar night. The altitude and phase of the sun and moon, which mainly contribute to the NSB, change over hours and days, rather than minutes.

During a trigger scan, the rate of a single, active trigger group is measured by counting the digital trigger output for a fixed trigger threshold during a \SI{200}{ms} interval. The trigger threshold is increased and the measurement is repeated until the rate is below \SI{50}{Hz}. This procedure is repeated until all trigger groups have a lower trigger rate.
During data-taking, the next-neighbor coincidence between the trigger groups reduces the trigger rate substantially below the expected event rate of a few Hertz. The threshold is kept constant between two trigger scans, which is typically a duration of six hours.
During trigger scans no operational parameters (see \autoref{sec:ops}) are logged.
\paragraph*{Dark noise runs.}
As stated in \autoref{sec:ops} the telescopes are covered during a dark noise run to reduce the amount of ambient light. The trigger threshold is set low enough so the full event readout is triggered by thermal noise from the SiPM. While the typical dark noise rate of the SiPMs during operation in the polar winter is a few kHz, the dead time limits the rate to a maximum of \SI{10}{Hz}.
The dark noise run (as well as the muon run) are used to calibrate the gain of the SiPMs \cite{Biland:2014fqa} for the roof and field telescope as the high NSB makes a gain calibration challenging during data-taking. In addition, the dark noise runs are used to monitor the electronic noise, which is dominated by the same high NSB during data-taking. 
\paragraph*{Muon runs.}
During a muon run the telescopes are also covered as in the case of the dark noise run. The goal is to measure the Cherenkov light emitted by a relativistic muon traveling through a solid PMMA cone of the camera.
The light is thereby mostly contained in a single pixel. The muons are identified by the high photon count ($\mathcal{O}(100)$ photons) in a singular pixel.

The trigger set-up is similar to the dark count run, with the exception of the trigger threshold to achieve a rate per trigger group \SI{\le1}{Hz}.
The average emitted number of photons is primarily driven by the geometry of the cone and as such the same for each pixel. The measured amount of light is then mainly driven by absorption and scattering in the cone as well as the glue, so the data can be used to quantify the systematic uncertainty of light yield of each pixel.
\paragraph*{Pedestal runs.}
The measured DC offset voltages of the ring buffer, called pedestals, are temperature and cell dependent.
To accommodate for this behavior a pedestal calibration run is taken. During this, the Target DAQ is triggered by an internal \SI{400}{Hz} trigger, while the camera is not powered.
The internal trigger is synchronized with the ring buffer.
A delay between trigger and readout is then used to shift through all of the possible cells in the ring buffer of the Target until all of the 4096 cells of the storage have multiple measurements of the offset voltage.
For the following data-taking each event is then pedestal corrected, meaning that for each used cell the average value from the pedestal run is subtracted.
\section{SADC Monitoring} \label{sec:SADC}

\begin{figure}[h]
    \centering
    \includegraphics[width=0.85\textwidth]{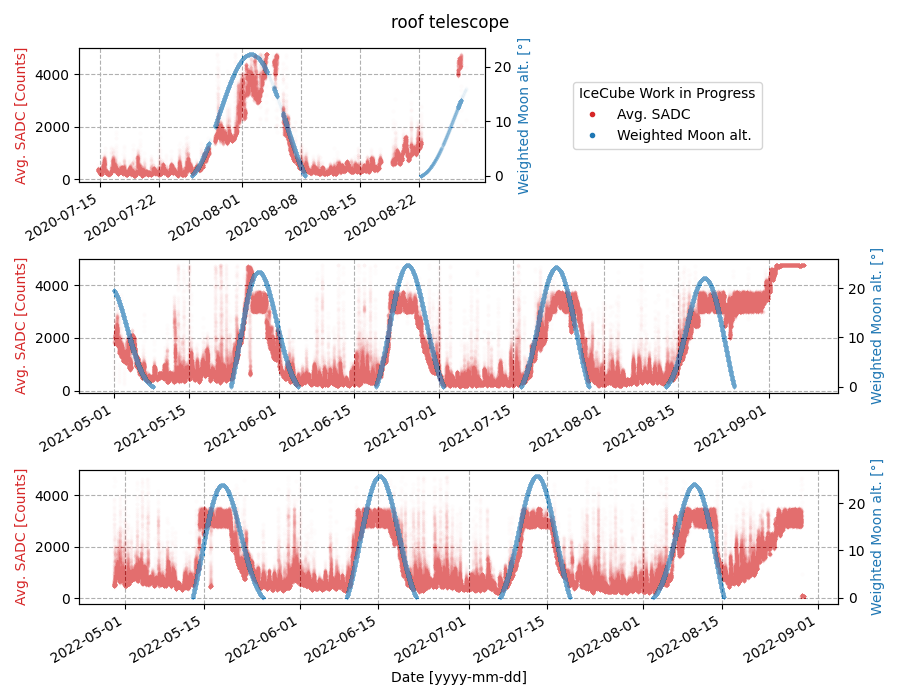}
    \caption{SADC measurements, averaged over the whole roof telescope. Also shown is the altitude of the moon, weighted with the percentage of the moon's illuminated surface area. The saturation-like effect is caused by an automated reduction of the SiPM gain. At the end of the season, the sunrise increases the NSB. The spikes between the moon phases are mostly caused by aurora australis.}
    \label{fig:MoonSADC}
\end{figure}

\begin{figure}[h]
    \centering
      \begin{subfigure}{\textwidth}
        \includegraphics[width=\textwidth]{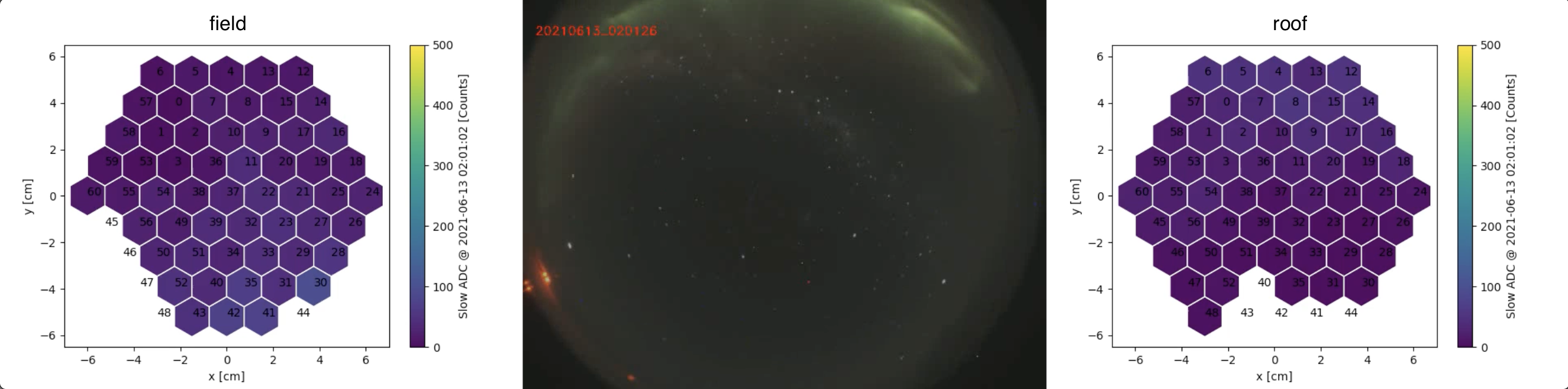}
        \caption{Low lightlevel}
    \end{subfigure}
     \begin{subfigure}{\textwidth}
        \includegraphics[width=\textwidth]{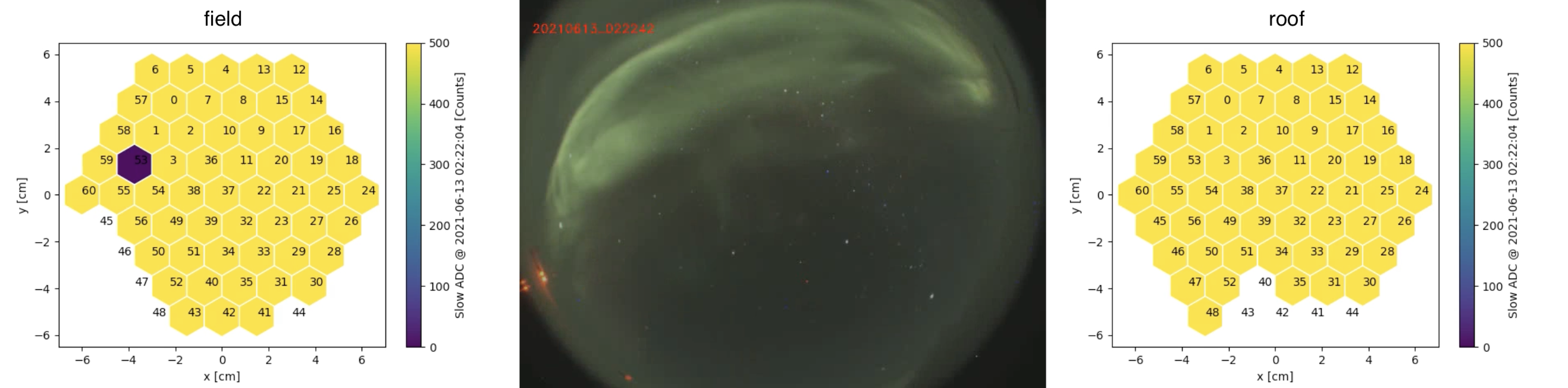}
        \caption{High lightlevel}
    \end{subfigure}
      
    \caption{
    Measurement of an Aurora with the telescope and the all-sky cam.
    On the left and right is a snapshot of the SADC values for each pixel of the field and roof telescope, respectively. The middle row shows a picture of the all-sky cam during the telescope measurements. For each pixel of the telescope, a median SADC value is subtracted, calculated from the first ten (aurora-free) minutes of the day. During low light levels, the upcoming aurora is visible at the border of the telescope. The missing pixels were not powered during data-taking. The SADC readout for pixel 53 of the field telescope is broken and therefore equal to 0.}
    \label{fig:SkycamADCpictures}
\end{figure}
The signal from each SiPM after pre-amplification and an additional gain stage is sent to a low pass filter and a consecutive operational amplifier, resulting in an integrated signal. It is digitized by a 16-bit ADC, the readout is also referred to as slow ADC (SADC).
The SADC measurements averaged over all active pixels can be found in \autoref{fig:MoonSADC}, as well as the altitude of the moon, weighted with the percentage of the illuminated, visible moon surface (e.g. 0\% is new moon, 100\% is full moon).
It can clearly be seen, that the moon has the strongest influence on the SADC measurements.

If the amount of light in the detector is too high for the dynamic range of the trigger threshold DAC the bias voltage of the SiPM and thus the gain is reduced by the software.
As a reduced gain also leads to a smaller integrated signal - a behavior similar to saturation during bright periods becomes visible. This can be seen during full moon, as well as at the end of the season of 2021 due to twilight.
The spikes between two moon phases are easily explained by aurora australis. In \autoref{fig:SkycamADCpictures} all-sky cam pictures taken during aurora activity can be seen.
The auroras are clearly visible in the SADC readout, showing that the SADC readout can be used as a quality cut for building a good run list.

The gathered data of the SADC can be used to improve future operations of the telescopes, similar as in \cite{Biland:2014fqa}. The trigger threshold is correlated to the amount of light in the detector. During future operations the SADC readout can be used as a swiftly measurable estimate of the amount of light in the detector and thus by proxy the necessary trigger threshold.

\begin{wrapfigure}{r}{0.6\textwidth}
    \centering
    \resizebox{0.56\textwidth}{!}{
    \input{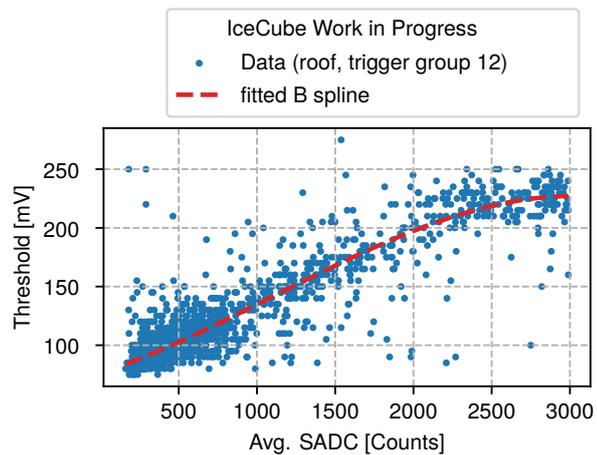}}
    \caption{Trigger threshold in dependence of the SADC readout. A B-spline was fitted to the data. The spline is planned to be used in future operations, to estimate the trigger threshold}
    \label{fig:SADC_Corr}
\end{wrapfigure}

For this purpose, it is necessary to derive an empirical relation between the amount of NSB and the set trigger threshold.
In \autoref{fig:SADC_Corr} the first SADC values after a trigger scan are plotted against the resulting trigger threshold of the trigger scan.
Exclusively 2021 was used, as the hourly trigger scans of the roof telescope lead to high statistics in regards to coinciding aurora activity, however, an inclusion of 2020 and 2022 data is possible.

As stated earlier in this section the SiPM supply voltage (and thus the gain) is reduced if the dynamic range of the trigger threshold DAC is not sufficient to accommodate for very high NSB. The data with reduced gain are not used for the analysis.
As stated in \autoref{sec:Calib} the readout of the SADC is currently not logged during a trigger scan. The resulting delay in SADC readout and trigger scan results in an additional variation between the set threshold and the measured SADC.
E.g., if the trigger scan is done shortly before an upcoming aurora, the result will be a low threshold and a delayed high SADC readout.
\newpage
To achieve a mean conversion from SADC to trigger threshold, a B-spline was fitted to the data with an outlier-robust regressor.
The resulting spline is planned to be used in the upcoming seasons to improve the trigger scan algorithm. Instead of scanning the whole range, the threshold is set to the value yielded by the B-Spline at a measured SADC.
If the resulting rate for the trigger group is too high, the trigger threshold can then be increased until the wanted rate is achieved and vice-versa.
This will reduce the amount of time during a trigger scan massively, enabling quick adjustment of the threshold during auroras, and shorter intervals between trigger scans while increasing the uptime of the telescopes.
\section{Outlook and Summary}

The two IceAct telescopes are operating successfully since 2020 in their current configuration. 
The method of operation during these three years is reported. The trigger threshold was tuned to achieve stable operations even during periods of high NSB.

\begin{wrapfigure}{R}{0.33\textwidth}
    \centering
    \includegraphics[width=0.3\textwidth]{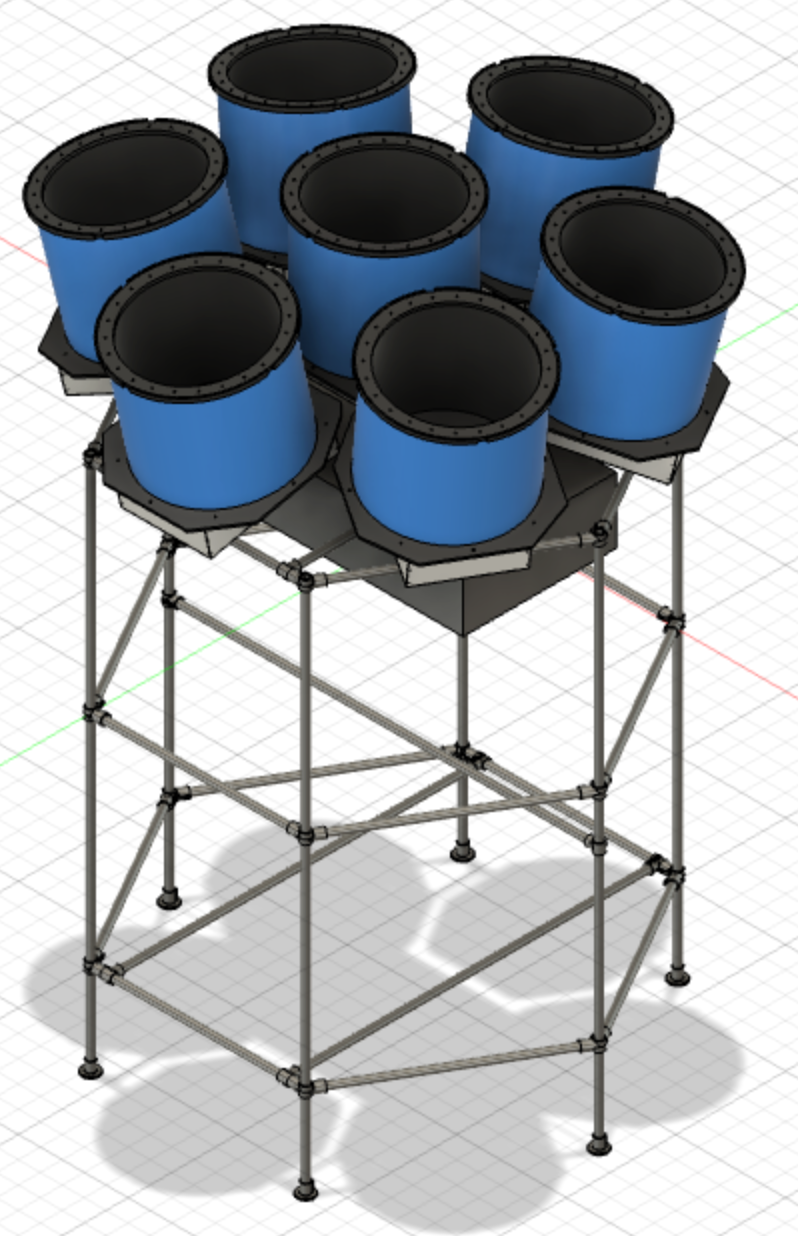}
    \caption{Rendering of a planned IceAct station.}
    \label{fig:CAD}
\end{wrapfigure}

A physics analysis based on the hybrid detection of cosmic rays is planned for the following years.
An updated simulation data set is currently in production for this purpose. Similar to the previously shown Monte-Carlo data set \cite{Schaufel:2019aef} it is a joined simulation of IceAct, IceTop, and the in-ice detector.
The atmospheric simulation was updated to improve the description of the absorption and scattering of Cherenkov photons within the south pole atmosphere. An array of 8 stations is simulated. More details of the simulated data set can be found in \cite{icrc23lpaul}.
An upgrade on the data-taking scripts is currently work-in-progress, with an implementation of the presented improved trigger scan algorithm, increasing the up-time of the telescopes. It is furthermore planned to store SADC measurements during trigger scans, to improve on the correlation shown in \autoref{fig:SADC_Corr}.

A multi-telescope setup of an IceAct station (see \autoref{fig:CAD}) is planned - configured in a so-called fly's eye configuration. This increases the field of view to \SI{36}{\degree}.
Six telescopes are built, four of them currently tested at the RWTH Aachen. The goal is a full station, plus the additional roof telescope for stereoscopic reconstructions of air showers.
\bibliographystyle{ICRC}
\bibliography{references}

%

\clearpage

\input{authorlist_IceCube.tex}

\end{document}

%% file: authorlist_IceCube.tex
\section*{Full Author List: IceCube Collaboration}

\scriptsize
\noindent
R. Abbasi$^{17}$,
M. Ackermann$^{63}$,
J. Adams$^{18}$,
S. K. Agarwalla$^{40,\: 64}$,
J. A. Aguilar$^{12}$,
M. Ahlers$^{22}$,
J.M. Alameddine$^{23}$,
N. M. Amin$^{44}$,
K. Andeen$^{42}$,
G. Anton$^{26}$
% ...
\\
\\
% ...
$^{12}$ Universit{\'e} Libre de Bruxelles, Science Faculty CP230, B-1050 Brussels, Belgium \\
% ...
$^{17}$ Department of Physics, Loyola University Chicago, Chicago, IL 60660, USA \\
$^{18}$ Dept. of Physics and Astronomy, University of Canterbury, Private Bag 4800, Christchurch, New Zealand \\
% ...
$^{22}$ Niels Bohr Institute, University of Copenhagen, DK-2100 Copenhagen, Denmark \\
$^{23}$ Dept. of Physics, TU Dortmund University, D-44221 Dortmund, Germany \\
% ...
$^{26}$ Erlangen Centre for Astroparticle Physics, Friedrich-Alexander-Universit{\"a}t Erlangen-N{\"u}rnberg, D-91058 Erlangen, Germany \\
% ...
$^{40}$ Dept. of Physics and Wisconsin IceCube Particle Astrophysics Center, University of Wisconsin{\textendash}Madison, Madison, WI 53706, USA \\
% ...
$^{42}$ Department of Physics, Marquette University, Milwaukee, WI, 53201, USA \\
% ...
$^{44}$ Bartol Research Institute and Dept. of Physics and Astronomy, University of Delaware, Newark, DE 19716, USA \\
% ...
$^{63}$ Deutsches Elektronen-Synchrotron DESY, Platanenallee 6, 15738 Zeuthen, Germany  \\
$^{64}$ Institute of Physics, Sachivalaya Marg, Sainik School Post, Bhubaneswar 751005, India \\
% ...

\subsection*{Acknowledgements}

\noindent
The authors gratefully acknowledge the support from the following agencies and institutions:
USA {\textendash} U.S. National Science Foundation-Office of Polar Programs,
U.S. National Science Foundation-Physics Division,
U.S. National Science Foundation-EPSCoR,
Wisconsin Alumni Research Foundation,
Center for High Throughput Computing (CHTC) at the University of Wisconsin{\textendash}Madison,
Open Science Grid (OSG),
Advanced Cyberinfrastructure Coordination Ecosystem: Services {\&} Support (ACCESS),
Frontera computing project at the Texas Advanced Computing Center,
U.S. Department of Energy-National Energy Research Scientific Computing Center,
Particle astrophysics research computing center at the University of Maryland,
Institute for Cyber-Enabled Research at Michigan State University,
and Astroparticle physics computational facility at Marquette University;
Belgium {\textendash} Funds for Scientific Research (FRS-FNRS and FWO),
FWO Odysseus and Big Science programmes,
and Belgian Federal Science Policy Office (Belspo);
Germany {\textendash} Bundesministerium f{\"u}r Bildung und Forschung (BMBF),
Deutsche Forschungsgemeinschaft (DFG),
Helmholtz Alliance for Astroparticle Physics (HAP),
Initiative and Networking Fund of the Helmholtz Association,
Deutsches Elektronen Synchrotron (DESY),
and High Performance Computing cluster of the RWTH Aachen;
Sweden {\textendash} Swedish Research Council,
Swedish Polar Research Secretariat,
Swedish National Infrastructure for Computing (SNIC),
and Knut and Alice Wallenberg Foundation;
European Union {\textendash} EGI Advanced Computing for research;
Australia {\textendash} Australian Research Council;
Canada {\textendash} Natural Sciences and Engineering Research Council of Canada,
Calcul Qu{\'e}bec, Compute Ontario, Canada Foundation for Innovation, WestGrid, and Compute Canada;
Denmark {\textendash} Villum Fonden, Carlsberg Foundation, and European Commission;
New Zealand {\textendash} Marsden Fund;
Japan {\textendash} Japan Society for Promotion of Science (JSPS)
and Institute for Global Prominent Research (IGPR) of Chiba University;
Korea {\textendash} National Research Foundation of Korea (NRF);
Switzerland {\textendash} Swiss National Science Foundation (SNSF);
United Kingdom {\textendash} Department of Physics, University of Oxford.